# High-field magnetoresistance of Fe/GaAs/Fe tunnel junctions

M. Zenger, J. Moser, W. Wegscheider and D. Weiss

*Experimentelle und Angewandte Physik, Universität Regensburg, 93040 Regensburg, Germany*

T. Dietl

*Institute of Physics, Polish Academy of Sciences, Al. Lotnikow 32/46, 02-668 Warszawa, Poland*

We investigate transport through 6 to 10 nm thin epitaxial GaAs(001) barriers sandwiched between polycrystalline iron films. Apart from a pronounced tunneling magnetoresistance effect (TMR) at low magnetic fields we observe a distinct negative magnetoresistance (MR) at low and a positive MR at higher temperatures. We show that the negative MR contribution is only observed for the ferromagnetic iron contacts but is absent if iron is replaced by copper or gold electrodes. Possible explanations of the negative MR involve suppression of spin-flip scattering or Zeeman splitting of the tunneling barrier.

Though impressive progress has been made in the field of magnetic semiconductors,[1] ferromagnetic metals are still prime candidates for electrical spin injection at room temperature. Given the 'conductivity mismatch problem'[2] effective spin injection from a ferromagnet (fm) into a semiconductor (sc) requires to insert adequately designed tunneling barriers at the fm/sc interface.[3,4] Schottky barriers, forming at most metal/sc junctions, offer a natural solution. Experiments in which spin injection is probed by analysing the degree of circularly polarized light in light emitting diode (LED) structures with epitaxially grown iron Schottky contacts indeed show spin injection[5,6] with a maximum polarization of up to 32 %.[7]

In order to probe the maximum achievable spin polarization in a transport experiment we fabricated GaAs tunnel junctions using a recently developed technique employing selective etching of AlGaAs sacrificial layers and thermal evaporation of iron contacts.[8,9] This results in a rectangular tunneling barrier, formed by the two Schottky contacts on both sides of the barrier.

To fabricate Fe/GaAs(001)/Fe tunneling elements we start from GaAs heterostructures, grown by molecular beam epitaxy containing a single crystalline GaAs tunnel barrier between two $Al_{0.72}Ga_{0.28}As$ sacrificial layers. By highly selective etching using hydrofluoric acid the sacrificial layers get removed and a polycrystalline metal contact is deposited. Before deposition of the second contact the original substrate is epoxy bonded upside down onto a s.i. GaAs(001) substrate. A schematic cross section of the completed device, shaped by optical lithography, is shown in the upper left inset of Fig. 1; a sketch of the top view with the 16 μm diameter tunneling junction in the center is shown in the bottom inset of Fig. 1.

To probe the high-field MR the device is mounted in a variable temperature insert (VTI) of a $^4$He-cryostat with a superconducting coil. The magnetic field was aligned in the plane of the tunneling barrier and the temperature in the VTI was varied between 4 K and 237 K in experiment. To probe the resistance drop across the barrier we measured in four-point configuration employing an HP Semiconductor Analyzer 4155A.

Our previous experiments provided convincing evidence that quantum mechanical tunneling is the dominating transport channel in our devices.[9] The tunneling current shows an almost perfect exponential dependence on barrier thickness d, $I \propto \exp(-2\kappa d)$. The damping constant $\kappa = 0.9 \pm 0.1$ nm$^{-1}$ is in close agreement with the model of Mavropoulos *et al.*[10] In the simplest approach the damping constant $\kappa$ is given by $\sqrt{2m\phi}/\hbar$ with m the effective electron mass in GaAs and the barrier height $\phi$ which is, for Schottky barriers on GaAs, about one half of the energy gap $E_g$, $\phi = E_g/2 = 0.76$ eV at 0 K. With this parameters $\kappa \approx 1.15$ nm$^{-1}$.

Fig. 1 displays the high-field MR of three Fe/GaAs/Fe tunnel junctions with 6, 8 and 10 nm thick GaAs barriers at 4.2 K. Their magnetic field dependence is strikingly different: while the MR for the sample with the thinnest barrier of 6 nm is negative over the investigated B-range, the MR for the thickest junction with 10 nm barrier becomes positive above ~ 5 T. The spike at B = 0 reflects antiparallel alignment of the iron contacts and is the TMR effect. Following recent work we used differently thick iron layers (4 nm and 20 nm) to establish different coercive fields of the two contacts at low temperatures.[9] The TMR effect obtained at a 8 nm thick barrier is shown on an expanded B-scale in Fig. 2. We observe a pronounced spin dependent resistance signal with a TMR ratio of 1,7% which is by a factor 8 higher than previously observed.[9] Here we used the usual definition that $TMR = (R_{\uparrow\downarrow} - R_{\uparrow\uparrow})/R_{\uparrow\uparrow}$, where $R_{\uparrow\uparrow}$ ($R_{\uparrow\downarrow}$) is the resistance for parallel (antiparallel) magnetization orientation. To extract the spin polarization of the current we used the Julliere model[11] which relates the TMR to the spin polarization P: $TMR = 2P^2/(1-P^2)$. This results in a spin polarization of about 9% for the data shown in Fig. 2. The fact that the spin polarization is significantly higher than previously reported we ascribe to new heterostructure material with reduced interface roughness and a reduction of the highest process temperature (epoxy cure) down to 80° C. Annealing at 120°C for 2 h decreases the TMR effect to 1.2% and indicates the sensitivity of the effect to layer intermixing or impurity diffusion.

Though our iron contacts are non epitaxial and though we probe transport across two Fe/GaAs interfaces our polarization values are comparable to values obtained on epitaxially grown Fe/AlGaAs LED devices.[6,7]

Now we return to the high-field MR displayed in Fig. 1. The observation of a negative MR surprises since one expects an increasing resistance with increasing in-plane field. In the simplest picture, the tunneling electrons, moving perpendicular to the applied field, follow a curved trajectory through the barrier. The longer path results in an increased resistance. Eaves *et al.*[12] obtain for the tunnel current $I(B) = I_0 \exp(-\beta B^2)$ with $\beta = e^2 d^3 \kappa / 6m\phi$. Here, $I_0$ is the current at B = 0. To test this prediction and to demonstrate that the negative MR is linked to the ferromagnetism of the contacts we fabricated reference Au/GaAs/Au and Cu/GaAs/Cu tunnel junctions. Corresponding data are shown in Figs. 3(a) and 3(b). As expected the positive MR increases with increasing barrier thickness, shown in Fig. 3(a). Though the expression for I(B) does not fit quantitatively, the dependence of I(B) on the barrier thickness d is qualitatively correct. However, the most important observation here, displayed in Fig 3(b), is that the tunneling current is independent of temperature. If we assume that the MR shown in Fig. 1 consists of two contributions, a temperature independent positive MR, due to the increased tunneling path, and an unknown one, associated with the ferromagnetic contacts, we can extract the latter. This is shown in Fig. 4. Fig. 4(a) displays the raw data showing a gradual increase of positive MR with increasing T. We then subtract from each trace the T-independent contribution, denoted as 'fit' in Fig. 4(a). We approximated this contribution by subtracting a linear contribution with slope dR(B, 237 K)/dB|$_{B\sim 0}$ from the R(B, 237 K) trace. Subtracting the resulting 'fit' from the raw data gives a linear negative MR displayed in Fig. 4(b). The slope of the negative MR as a function of T is shown for two samples in the inset of Fig. 4(b). We ascribe this negative MR to the spin polarization of the tunneling current.

To explain the linear negative MR we considered (i) suppression of spin-flip scattering at the interface and at magnetic impurities in the barrier or (ii) spin-dependent barrier heights due to the Zeeman effect. The spin splitting of the conduction band edge causes different tunneling barriers for spin-up and spin-down electrons. Assuming independent spin channels the difference $\frac{\Delta\sigma_{\uparrow,\downarrow}}{\sigma} = \frac{\sigma_{\uparrow,\downarrow} - \sigma(B=0)}{\sigma(B=0)}$ for spin-up ($\sigma_\uparrow$) and spin-down ($\sigma_\downarrow$) conductivities, can be approximated by $\frac{\Delta\sigma}{\sigma} \approx \pm\frac{\kappa d}{2\phi}g^*\mu_B B$ with $\kappa = \sqrt{2m\phi}/\hbar$. For $\phi = 0.76$ eV, $g^* = -0.44$, $\kappa = 1.15$ nm$^{-1}$, $m = 0.067 m_0$ with $m_0$ the free electron mass, d = 8 nm and B = 10 T we obtain for the spin-up channel $\Delta\sigma/\sigma \approx +0.15\%$ and for spin-down $\Delta\sigma/\sigma \approx -0.15\%$. While our ansatz gives a linear MR with an upper estimate of $\Delta\sigma/\sigma \approx \pm 0.15\%$ (obtained for the unrealistic assumption of a half metallic contact with 100% spin polarization), the effect is by a factor 6-7 smaller than the resistance change observed in experiment (cf. 4 K-trace in Fig. 4(b)). By using a three band model[13] $\Delta\sigma/\sigma$ becomes of the order of 1%. However, the temperature dependence of $\Delta\sigma/\sigma$ cannot be explained by our simple tunneling picture. Furthermore the sign of the MR effect would be in conflict with recent experiments claiming that the tunneling current is carried by the iron's majority spins, i.e. by spins which are aligned antiparallel to the applied magnetic field.[14] Since the g-factor in GaAs is negative the tunneling barrier increases for spin-down electrons (antiparallel to the applied field) while getting lowered for spin-up electrons. Contrary to our experiment this should cause a linearly increasing MR.

Spin-flip scattering at the interface or in the barrier might be an alternative source of the negative MR. The diffusion of iron 'dopes' the GaAs barrier with spin-flip scatterers. ESR experiments showed that iron exists in GaAs as $Fe^{3+}$-ions with 3d$^5$-electron configuration[15] having, in the presence of B, Zeeman split levels of spin $S_z$ = ±1/2, ±3/2, ± 5/2. At low T spin-flip scattering is expected to get suppressed with increasing B since electrons condense in the ground state $m_j$ = -5/2. Therefore a decreasing resistance, as in experiment, is expected for sufficiently low bias. With increasing temperature and hence finite occupancy of higher spin states spin-flip scattering gets reinstalled. The T-dependence of the MR should hence be con-nected to the occupancy 1-$\vartheta$ of the excited spin levels. In the inset of Fig. 4(b) the ground state occupancy $\vartheta$ at 10 T is plotted as a function of T. While the occupancy nearly saturates above 50 K the slope of the negative MR decreases over the entire temperature range. Hence spin-flip scattering in the barrier alone does not account for our observations.

In summary we have observed a linear negative high-field MR in Fe/GaAs/Fe tunnel elements. Comparison of our results with simple models involving spin-dependent tunneling barriers and spin-flip scattering in the barrier suggest that additional effects like magnetic disorder at the interface and/or Zeeman splitting of surface states must be considered in a more detailed theory.

Financial support by the BMBF (grant 01BM921) and by the DFG (Forschergruppe 370) is gratefully acknowledged. T.D.'s research in Germany is supported by the Alexander von Humboldt foundation. The authors thank Gerrit Bauer, Jens Siewert and Wulf Wulfhekel for discussions.

# Figure Captions

Fig. 1: High-field MR measured at 4.2 K and a bias voltage of 20 mV for 6 nm, 8 nm and 10 nm thin GaAs barriers, sandwiched between two iron contacts. The sharp peak at B ~ 0 is the TMR effect. The upper inset shows the cross section, the lower one a schematic top view of the tunnel junction.

Fig. 2: TMR traces of a tunnel junction with d = 8 nm at 4.2 K measured at a bias voltage of 5 mV.

Fig. 3: Tunneling resistance of junctions with nonmagnetic contacts at 4.2 K. (a) Dependence of positive MR on barrier thickness d for Au/GaAs/Au junctions. The solid lines correspond to the model of Eaves et al.[12] fitted to the 6 nm trace. (b) Temperature dependence of the normalized resistance of a Cu/GaAs/Cu junction with d = 10 nm. The bias voltage was in both cases 20 mV.

Fig. 4: (a) Normalized resistance of a Fe/GaAs/Fe junction with d = 8 nm as a function of temperature. The fit of the positive MR contribution is obtained by taking the difference between the 237 K trace and a line with slope $dR(237\,K)/dB$ taken close to B = 0 outside the TMR regime. (b) After subtraction of the T independent positive MR contribution ('fit') from all traces in (a) a linear negative MR is obtained. The temperature dependence of the slope of the linear MR is shown in the inset for samples with 6 nm and 8 nm barrier thickness. The dashed line shows the temperature dependence of the ground state occupancy $\vartheta$ of a $Fe^{3+}$ impurity in GaAs.

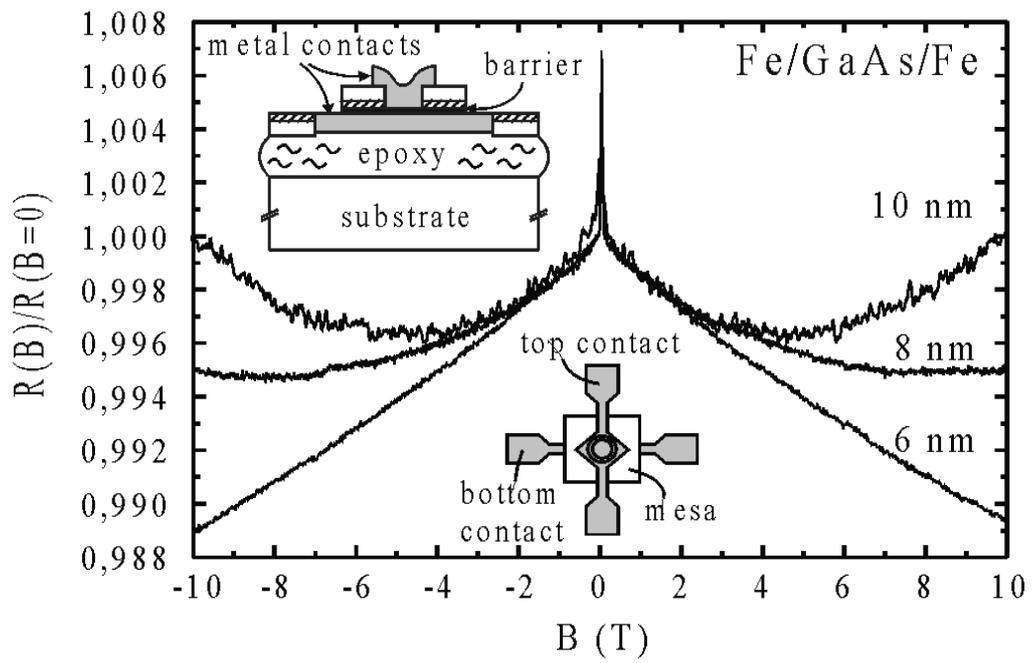

**Fig. 1/4**

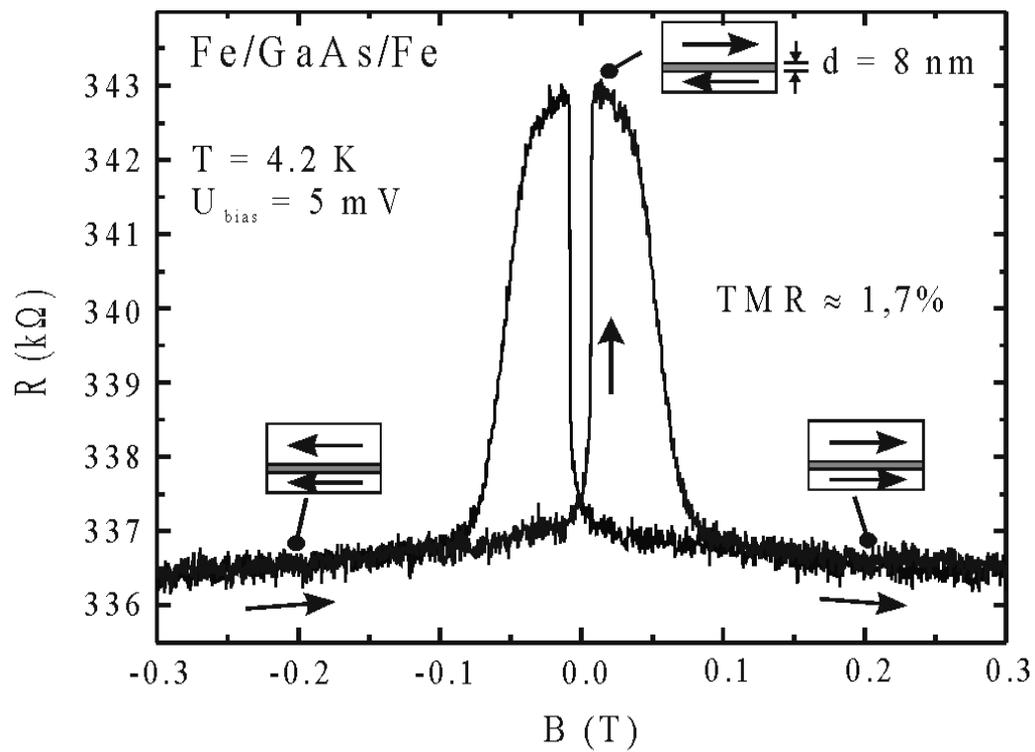

Fig. 2/4

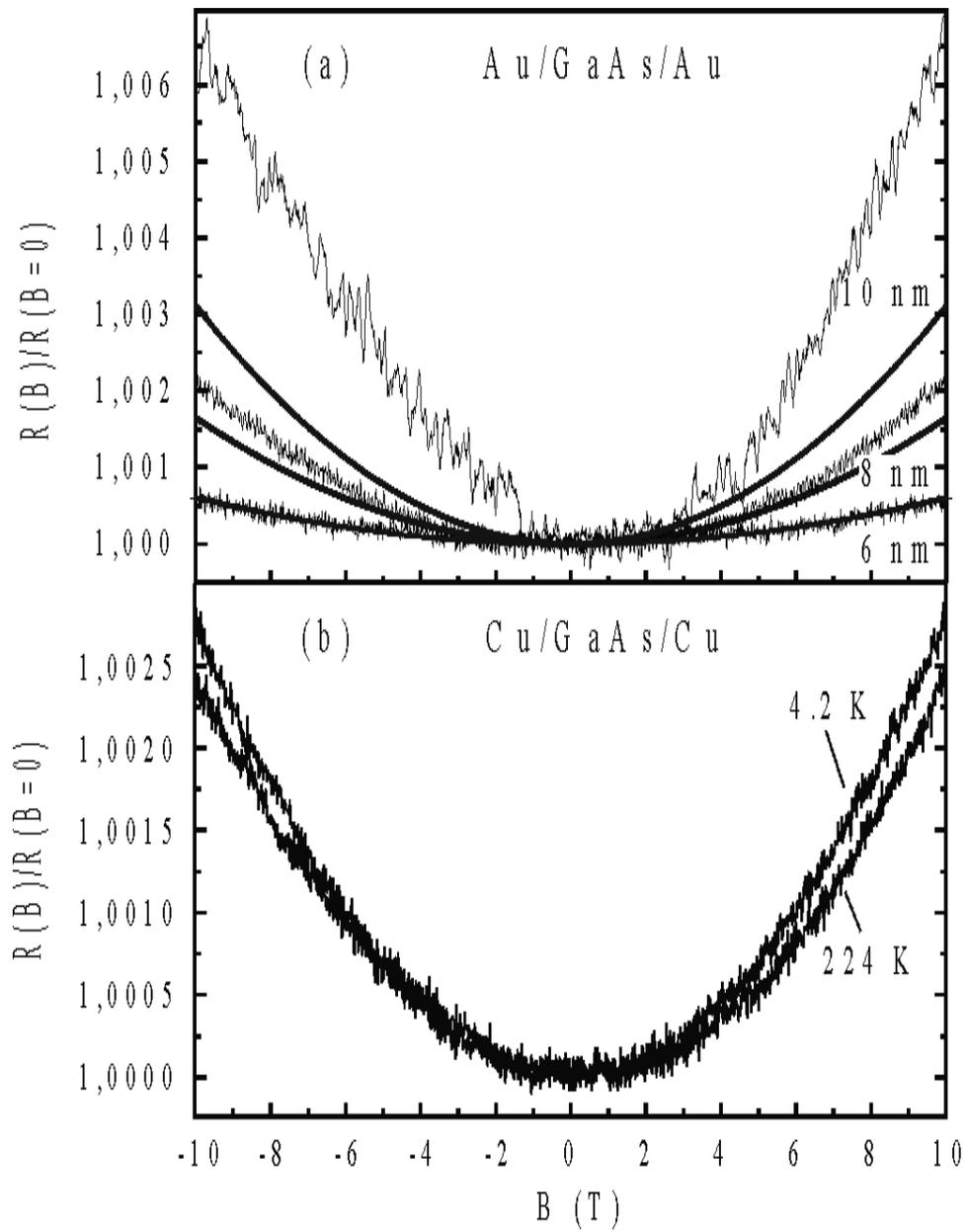

Fig. 3/4

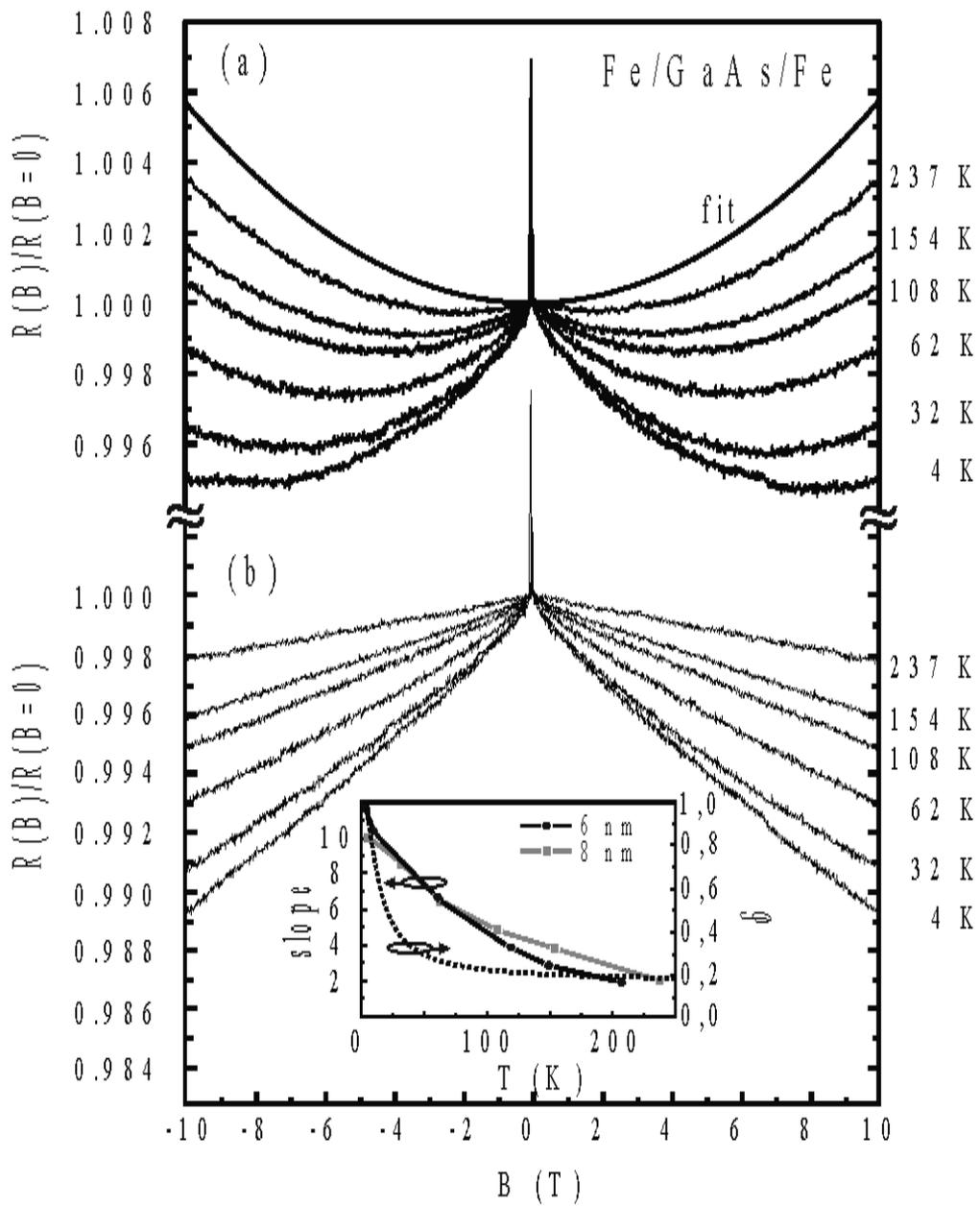

Fig. 4/4